# Discovery and characterization of magnetism in sigma-phase intermetallic Fe-Re compounds

J. Cieślak[1*], S. M. Dubiel[1], M. Reissner[2] and J. Tobola[1]

[1]AGH University of Science and Technology, Faculty of Physics and Applied Computer Science, PL-30-059 Krakow, Poland

[2]Institute of Solid State Physics, Vienna University of Technology, A-1040 Wien, Austria

Systematic experimental (vibrating sample magnetometry) and theoretical (electronic structure calculations using charge and spin self-consistent Korringa-Kohn-Rostoker Green's function method) studies were performed on a series of intermetallic sigma-phase $Fe_{100-x}Re_x$ ($x$ = 43-53) compounds. Clear evidence was found that all investigated samples exhibit magnetism with an ordering temperature ranging between ~65 K for $x$ = 43 and ~23 K for $x$ = 53. The magnetism was revealed to be itinerant and identified as a spin-glass (SG) possibly having a re-entrant character. The SG was found to be heterogeneous viz. two regimes could be distinguished as far as irreversibility in temperature dependence of magnetization is concerned: (1) of a weak irreversibility and (2) of a strong one. According to the theoretical calculations the main contribution to the magnetism comes from Fe atoms occupying all five sub lattices. Re atoms have rather small moments. However, the calculated average magnetic moments are highly (ferromagnetic ordering model) or moderately (antiparallel ordering model) overestimated relative to the experimental data.

[*]Corresponding author: Cieslak@fis.agh.edu.pl

## 1. Introduction

The sigma (σ) phase (space group $D_{4h}^{14}$ - $P4_2/mnm$) is possibly the best-known and most frequently investigated example of Frank-Kasper (FK) phases that occur in alloys in which at least one element is a transition metal. Their characteristic features are high coordination numbers (12-16) and lack of stoichiometry. The originating therefrom topological and chemical complexity as well as the diversity of physical properties that can be tailored by changing constituting elements and/or their relative concentration, make them an attractive yet challenging subject for investigation. The latter is further prompted by the deteriorating effect the FK-phases have on materials in which they precipitate, although attempts have been undertaken to profit from their high hardness and use their precipitation for strengthening purposes e. g. [1,2].

Concerning magnetic properties of σ in binary alloys, so far only σ in Fe-Cr and Fe-V systems was definitely revealed to be magnetic [3-5]. However, its magnetism was shown to be more complex than initially regarded viz. in both cases it has a re-entrant character [6]. Furthermore, NMR measurements performed on σ-FeV samples gave evidence that vanadium atoms present in all five sub lattices are magnetic [7].

Here, based on a systematic experimental (vibrating sample magnetometry, ac susceptibility) and theoretical (electronic structure calculations using charge and spin self-consistent Korringa-Kohn-Rostoker (KKR) Green's function method as well as KKR combined with the coherent potential approximation (CPA) technique) studies a clear evidence is reported in favor of the magnetic properties of σ in a series of Fe-Re compounds.

## 2. Experimental

### 2. 1. Samples

A series of σ-$Fe_{100-x}Re_x$ ($x$ = 43, 45, 47, 49, 52.6 and 53) was prepared by the following way: powders of elemental iron (3N+ purity) and rhenium (4N purity) were mixed in appropriate proportions and masses (2 g), and next pressed to pellets. The pellets were subsequently melted in an arc furnace under protective atmosphere of argon. The produced ingots were re-melted three times to enhance the degree of homogeneity. Next, they were vacuum annealed at 1803 K for 5 hours and, finally, quenched into liquid nitrogen. The mass loses of the fabricated samples were less than 0.01% of their initial values, so it is reasonable to take their nominal compositions as real ones. X-ray diffraction patterns recorded on the powdered samples gave evidence that their crystallographic structure was σ. More details on structural and electronic properties of σ in the studied samples can be found elsewhere [8].

### 2. 2. Magnetic measurements

Dc magnetic measurements were performed in a 9 T Physical Properties Measurement System (PPMS) with Vibrating Sample Magnetometer option. Temperature dependence of magnetization was measured under both zero-field cooling (ZFC) and field-cooling (FC) conditions in temperature range 3 to 300 K. Used field was 10 mT. To avoid influence of remanence of superconducting coil and that of earth field, in advance to each measurement field was set exactly to zero with help of a compensation coil. Magnetization curves versus magnetic field were recorded at various temperatures in the range from 4 to 300 K in fields up to 9 T, using field sweep rate of 20 mT/s.

## 3. Theoretical calculations

The charge and spin self-consistent Korringa–Kohn–Rostoker (KKR) Green's function method [9-11] was used to calculate the electronic structure of the studied samples. The crystal potential of muffin-tin (MT) form was constructed within the local density approximation (LDA) framework using the Barth–Hedin formula [12] for the exchange–correlation part. The group symmetry of the unit cell of the σ-phase was lowered to a simple tetragonal one to allow for various configurations of Fe/Re atoms. The experimental values of lattice constants and atomic positions were used in all computations [8]. For fully converged crystal potentials electronic density of states (DOS), total, site-composed and $l$-decomposed DOS (with $l_{max}$ = 2 for Fe and Re atoms) were derived. Fully converged results were obtained for ~120 special $k$-point grids in the irreducible part of the Brillouin zone but they were also checked for convergence using a denser $k$-mesh. Electronic DOS were computed using the tetrahedron $k$-space integration technique and ~700 small tetrahedral. The KKR calculations were carried out for 17 unit cells with different configurations of Fe and Re atoms on all five sub lattices, keeping experimental lattice constants in all cases as constraints. The configurations were chosen in such a way that each possible number of Fe atoms being the nearest-neighbors for a given lattice site, $NN_{Fe}$, had been taken into account. More details relevant to the calculations protocol can be found elsewhere [13, 14].

In order to study the magnetism of the sigma-FeRe alloy spin polarized KKR as well as KKR-CPA calculations have been performed. A collinear ordering of magnetic moments on Fe and Re atoms, as allowed by the used computer code, were assumed in a ferromagnetic (FM) state both in the ordered approximants and in the disordered alloys. Parallel a relatively complex model of the magnetic structure, based on antiparallel alignments of magnetic moments within selected sub lattices called APM, was applied (spin-polarized KKR-CPA calculations were also performed for such magnetic unit cell).

## 4. Results and discussion

### 4. 1. Magnetization vs. temperature

FC and ZFC magnetization curves recorded on all samples displayed in Fig. 1 give a clear evidence that all samples (a) are magnetic at temperatures less than ~80 K, and (b) show irreversibility. Such behavior is a characteristic feature of spin-glasses and it was also recently revealed for the σ-phase Fe-Cr and Fe-V intermetallic compounds [6]. The magnetic ordering temperature, $T_C$ (susceptibility measurements - discussed below - gave evidence that the magnetic state orders ferromagnetically) was determined from the high temperature side of $M(T)$ curves in two ways: (1) from the point of inflection and (2) from extrapolation of the *FC*-curves to $M = 0$. As shown in Fig. 2 the two sets of $T_C$-values fit well with each other. It is also evident that the $T_C$ decreases with the increase of rhenium content, $x$, from ~65 K at $x$ = 43 to ~20 K at $x$ = 53.

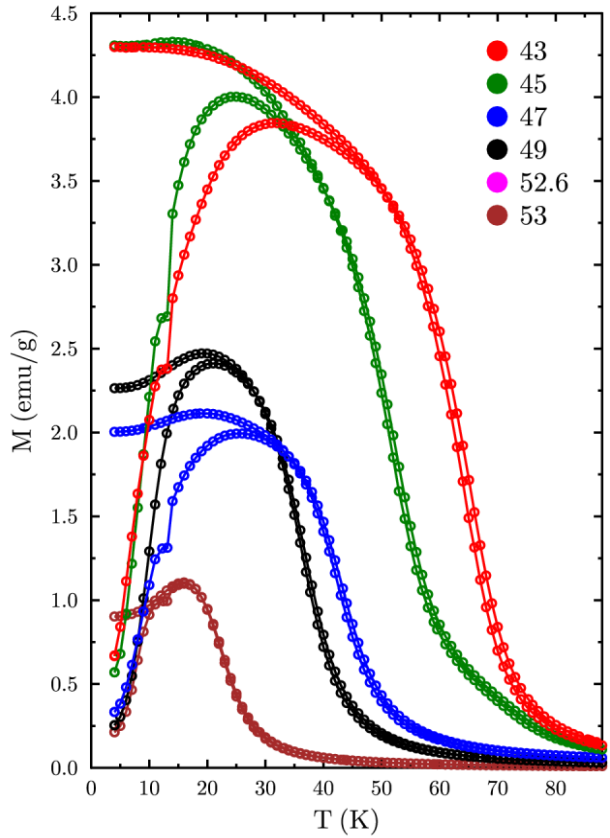

Fig. 1(Color online) Field cooled and zero-field cooled magnetization curves of the studied samples recorded in external magnetic field of 10 mT. Figures in the legend stay for the content of rhenium in at%.

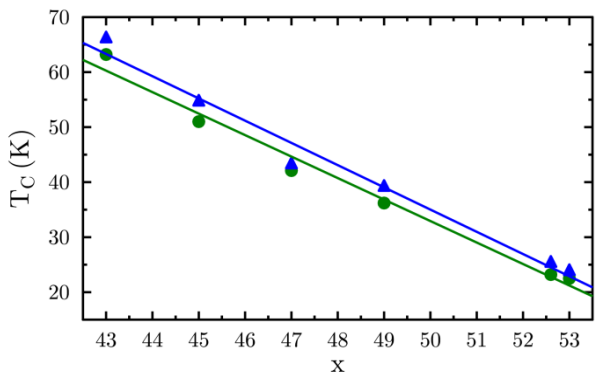

Fig. 2 (Color online) Magnetic ordering temperature, $T_C$, versus Re-content, $x$, as derived from the point of inflection (circles), and from an extrapolation to zero of the $M_{FC}(T)$ curves (triangles). Solid lines represent the linear best-fits to the data.

The temperature, at which the FC and ZFC curves go apart from each other marks the transition into a spin glass (SG) state, and it is known as a spin-freezing temperature, $T_f$. Below this temperature irreversibility effects characteristic of SG sets in. Another temperature, $T_{si}$, that has been often used in description of spin-glasses is the one at which the ZFC-curve has a maximum. This temperature is interpreted as indicative of the onset of strong irreversibility effects as opposed to weak ones that occur in the range between $T_f$ and $T_{si}$. The dependence of both $T_f$ and $T_{si}$ on $x$ for the studied samples is presented in Fig. 3.

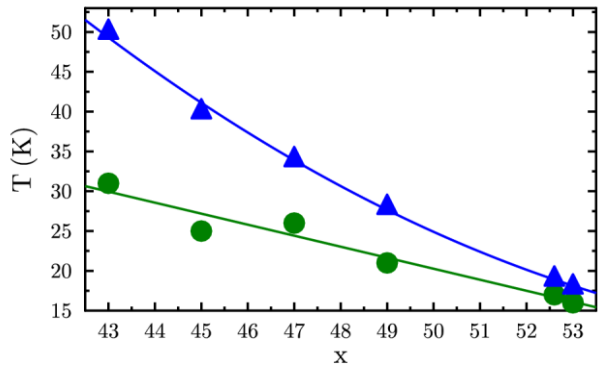

Fig. 3

(Color online) Dependence of the spin-freezing temperature $T_f$ (triangles) and that of the strong irreversibility temperature $T_{si}$ (circles) on rhenium content, $x$. The lines represent a guide to the eye.

One can see that both temperatures decrease quite monotonically with $x$, and eventually, they meet at $x \approx 53$. Consequently, the weak irreversibility region of the SG defined by the two temperatures shrinks with $x$. Based on the data shown in Figs. 2 and 3 a magnetic phase diagram of the σ-phase Fe-Re alloy system could be outlined – see Fig. 4.

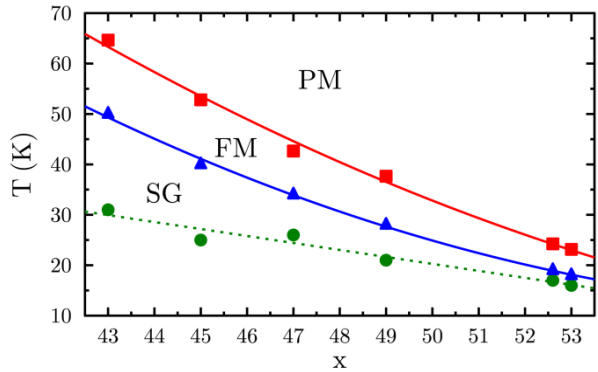

Fig. 4

(Color online) Magnetic phase diagram derived from $M(T)$ measurements. Paramagnetic state is indicated by PM, ferromagnetic-like one by FM, and spin-glass one by RSG. The line marking the transition into the strong irreversibility range of SG is indicated by the dotted line.

From the $M(T)$-curves an information on a degree of reversibility of the SG state as a function of temperature can be obtained. For that purpose we calculated a relative difference between $M_{FC}(T)$ and $M_{ZFC}(T)$ curves and present it in Fig. 5. It is clear that the degree of irreversibility grows with decreasing temperature. The growth is firstly slow, which reflects a weak irreversibility regime, and next it proceeds much faster indicating a strong irreversibility regime.

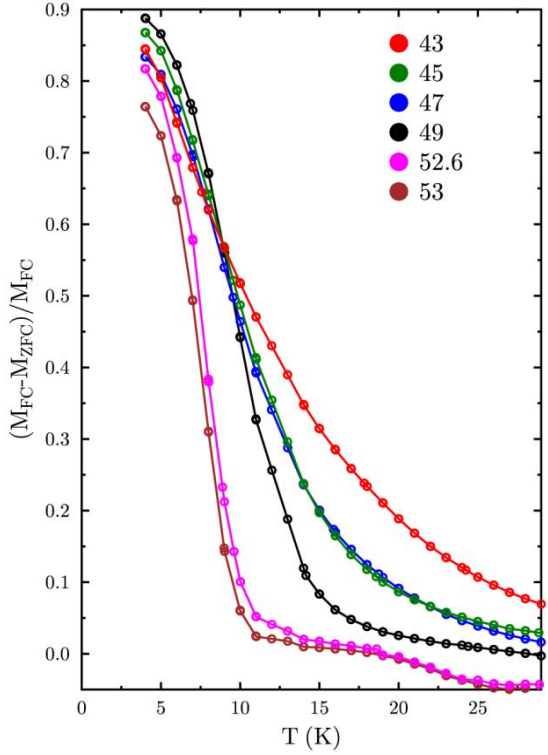

Fig. 5

(Color online) Temperature dependence of the relative difference in area under the $M_{FC}(T)$ and $M_{ZFC}(T)$ curves shown in Fig. 1. The legend shows the rhenium content in at%.

To get an information on the average degree of irreversibility for a given sample, we calculated an area under the $M_{FC}(T)$ curves, $A_{FC}$, from the equation:

$$A_{FC}(x) = \int_{4.2}^{90} M_{FC}(x,T) dT \qquad (1)$$

And under the $M_{ZFC}$ curves, $A_{ZFC}$, from the equation:

$$A_{ZFC}(x) = \int_{4.2}^{90} M_{ZFC}(x,T) dT \qquad (2)$$

And next we calculated a relative difference between them, $\Delta A = 100(A_{FC}-A_{ZFC})/A_{FC}$, for each value of $x$. This quantity can be regarded as a measure of the degree of irreversibility. Data obtained in this way are displayed in Fig. 6.

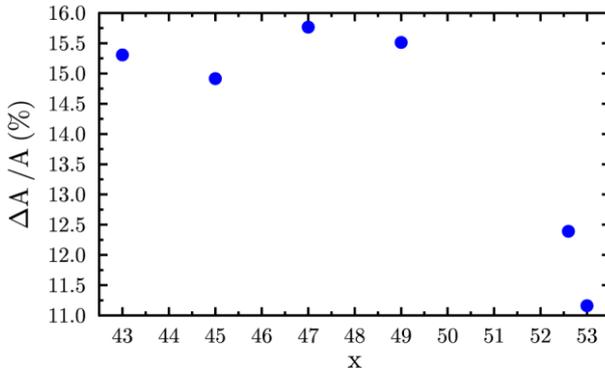

Fig. 6

(Color online) A relative degree of irreversibility, $\Delta A/A$, versus Re content, $x$, for the studied samples.

It can be easily noticed that the for $43 \leq x \leq 49$, $\Delta A/A$, hardly depends on $x$, while for the two Re-most concentrated samples it has lower values. The latter samples have also the weakest magnetic properties, as shown in Figs. 4 and 10, and the smallest differences between the magnetic ordering temperature and the spin-freezing one.

4.2. AC susceptibility

A cusp in the ac susceptibility, $\chi$, and its frequency dependence have been regarded as the most prominent signature of the spin-glasses. Having this in mind, corresponding measurements were performed on the $\sigma$-Fe$_{47}$Re$_{53}$ sample for the frequency ranging between 8 and 1000 Hz. The results shown in Fig. 7 give a clear evidence that $\chi$ has a characteristic cusp that shifts to higher temperature values as frequency grows. The figure of merit i.e. a relative shift of the spin-freezing temperature, defined by the cusp, per a decade of frequency has a value of 0.008 which is typical of metallic spin-glasses with long-range interactions between magnetic moments (Ruderman-Kittel-Kasuya-Yosida).

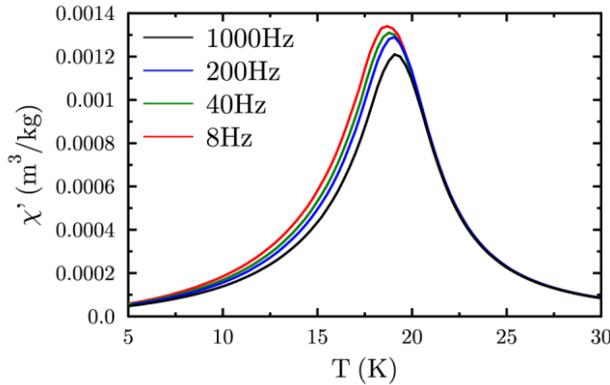

Fig. 7 Real part of the ac susceptibility, $\chi'$, measured on the $\sigma$-$Fe_{47}Re_{53}$ sample.

## 4. 3. Magnetization vs. magnetic field

To get more insight into the magnetism of the investigated $\sigma$-FeRe compounds magnetization measurements in external magnetic field, $\mu_oH$, were performed. The obtained curves are presented in Fig. 8a and 8b. It is evident that even at 9 T the saturation state has not been achieved. The latter is in line with the finding that it is the spin-glass that constitutes the magnetic ground state of the investigated samples.

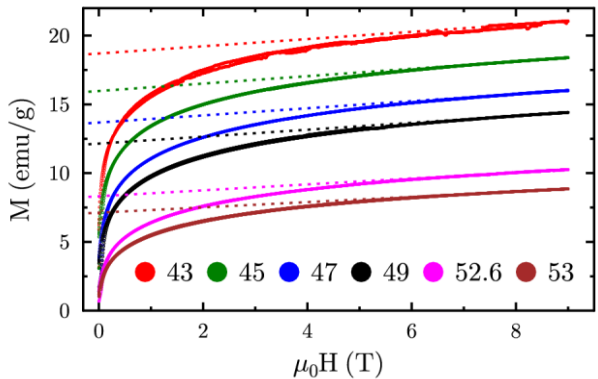

Fig. 8a

(Color online) Magnetization curves measured at 4.2K as a function of an external magnetic field, $\mu_oH$. Dotted lines represent extrapolation of high-field linear parts of the *M*-curves. The legend marks the content of Re.

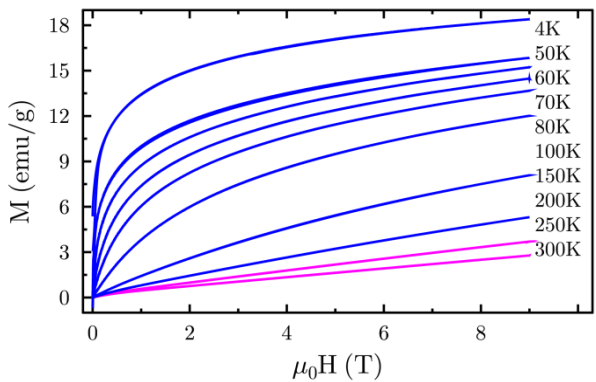

Fig. 8b

(Color online) Magnetization curves measured on the $\sigma$-$Fe_{55}Re_{45}$ alloy at different temperatures (indicated in K) as a function of an external magnetic field, $\mu_oH$. The straight lines were obtained for T=250 and 300K.

Two information relevant to the magnetism of the $\sigma$-phase Fe-Re compounds can be derived from the *M(T)* and *M(H)* measurements. First, one can conclude on the kind of magnetic ordering (ferromagnetic or antiferromagnetic). The conclusion can be drawn from the low-field susceptibility. The linear part of the reciprocal low-field dc magnetic susceptibility versus temperature is shown in Fig. 9 for all *x*-values. The extrapolations to zero of the linear fragments intersect the temperature axis on its positive side, proving thereby that the paramagnetic Curie temperatures (Weiss constants), Θ, are positive i.e. a coupling of the magnetic moments has a ferromagnetic-like character. However, as evidenced in Fig. 8b, the *M(H)*-curves measured at different temperatures show curvature, which is normally regarded as a sign of a ferromagnetic coupling, at temperatures as high as 200 K which is much higher than the corresponding paramagnetic Curie temperature (the latter 10-15K higher than the corresponding $T_C$-values) as well as the temperature determined by the inflection point of *M(H)* and denoted above as $T_C$. However, there is also experimental evidence that such curvature in *M(H)* measurements was observed in Cr+2.7%Fe$_{100-x}$V$_x$ alloys in a paramagnetic state i.e. $x \geq 5$ [21]. Both systems i.e. σ-FeRe and Cr-Fe-V have in common an itinerant character of magnetism. Whether or not the presently determined $T_C$-line represents the real magnetic ordering temperature cannot be uniquely decided based on the present experiments. Neutron diffraction measurements which are planned to be carried out will hopefully deliver the solution.

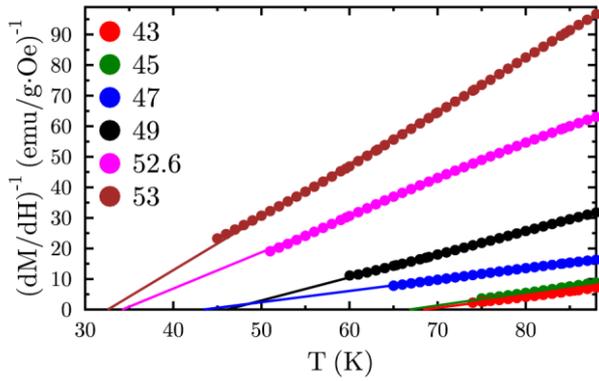

Fig. 9

(Color online) Linear parts of the reciprocal low-field dc susceptibility and their extrapolations to zero value for the investigated samples versus temperature, *T*. The legend shows the concentration of rhenium.

Second, by extrapolating the high-field linear part of the *M(H)*-curves to zero field, an average magnetic moment per atom, μ, is estimated and presented in Fig. 10. It can be seen that μ is a monotonous function of *x* and its values range between ~0.4 μ$_B$ at *x* = 43 and ~0.2 μ$_B$ at *x* = 53. In other words, an increase of rhenium content causes weakening of magnetism of the investigated alloys. Interestingly, the values of <μ> are the highest in comparison to those determined for the σ-phase Fe-V and Fe-Cr compounds in a similar range of composition [5].

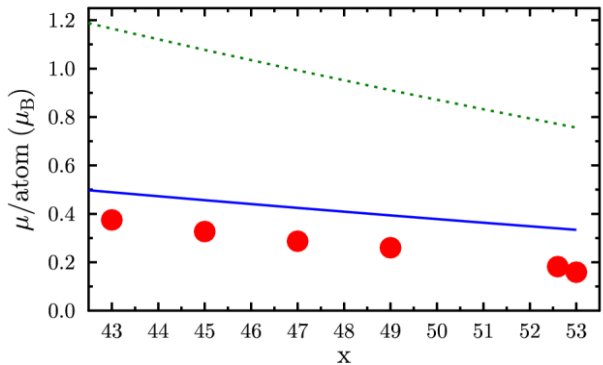

Fig. 10

(Color online) Average magnetic moment per atom, μ, as estimated from the in-field magnetization measurements (circles), and that calculated based on the FM model (dotted line) and on the APM model (solid line).

## 4. 3. Theoretical calculations

### 4.3.1. KKR calculations

The calculations outlined in section 3 yielded magnetic moments of Fe and Re atoms for all considered atomic configurations in each of the five sub lattices. The results obtained with the FM model can be seen in Fig. 11. For each of the five sub lattices there is a cloud of data representing magnetic moments of particular atomic configurations taken into account in the calculations. The Fe-site µ-values are positive for nearly all configurations and sub lattices, and for every sub lattice there is an increase of µ with the number of the nearest-neighbor Fe atoms, $NN_{Fe}$. The maximum calculated value of µ ranges between ~$2\mu_B$ for the site A and ~$3\mu_B$ for the other four sites. Based on these data, the average magnetic moment for each sub lattice, $<\mu>_k$, (k = A, B, C, D, E) was calculated. Its values are indicated in Fig. 11 by horizontal lines. Here, the highest value of ~2.3 $\mu_B$ represent Fe-atoms residing on site C, and the lowest one of ~0.9 $\mu_B$ those that occupy sites A.

Concerning the Re-site magnetic moments, they are significantly smaller, as shown in the lower panel of Fig. 12, and their values for particular configurations oscillate around zero, the maximum amplitude being that for the site B viz. between ±0.15 $\mu_B$. Also in this case the average µ-values were determined and presented as horizontal lines. One can see that in this case the average moments on Re-atoms residing on sites C and E are zero, while those occupying sites B and D have weakly negative values i.e. -0.02 $\mu_B$ for site A and -0.04 $\mu_B$ for site D. Knowing the $<\mu>_k$ – values for both elements and all sub lattices and the atomic occupancies of the latter as found elsewhere [8], the average magnetic moment per atom (Re or Fe), $<\mu>$, was computed.

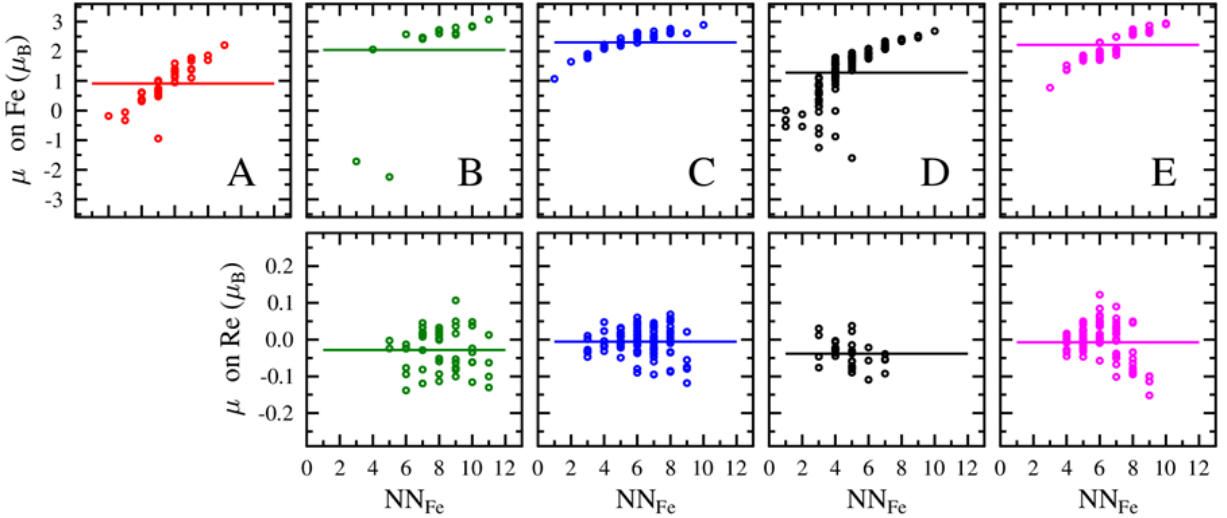

Fig. 11

(Color online) Theoretically calculated, using the FM model, magnetic moments on Fe and Re atoms occupying the five different sub lattices A, B, C, D and E versus number of nearest-neighbor Fe atoms, $NN_{Fe}$. The cloud of data seen for each sub lattice represents magnetic moments of different atomic configurations taken into account. Horizontal lines indicate the average values of the moments.

Its dependence on *x* is shown as a dotted line in Fig. 10. Although it correctly reproduces the experimental concentration dependence of μ, yet the absolute values are overestimated by a factor of over 2. This failure of the FM approach was also faced in the case of similar calculations performed for the other two intermetallic σ-phase compounds that exhibit magnetism viz. Fe-Cr [14-16] and Fe-V systems [17]. Consequently, based on group theory and a symmetry analysis, it was shown that an antiparallel ordering on C and D sub lattices could be possible. Therefore a model (APM approach) which allows antiparallel alignment on sites C and D was applied [16]. This model resulted in a significant reduction of <μ>, hence a better agreement with the experimental values was achieved. The results of the APM approach applied in the present case can be seen in Fig. 12. The main difference, as compared to the results obtained with the FM model (Fig. 11), concerns the sites C and D for which the clouds have separated into two branches: one corresponding to positive values of μ, the other one to negative ones. The average μ-values were calculated for all sub lattices and both atoms, and they are indicated in Fig. 12 as horizontal solid lines. The average values for the positive and negative clouds (sub lattices C and D) are marked by horizontal dotted lines. The $<\mu>_k$-values found with the APM model are strongly reduced relative to the corresponding ones presented in Fig. 12, what obviously results in a reduced average moment per atom. The concentration dependence of the latter is presented as a solid line in Fig. 10. A reduction of the discrepancy between the experimental data and the dotted line is clear, albeit the agreement is still unsatisfying even taking into account that the experimental points are underestimated due to the unsaturated nature of the *M(H)* curves. It seems that the main reason for the observed disaccord between the experimental and theoretical results is the co-linear magnetic ordering assumed in the calculations. The experimental findings reported in this paper give a clear evidence that in the ground state the ordering is not co-linear. Nevertheless, it seems that the APM approach is a step forward in a theoretical description of the σ-phase magnetism as it describes it as a mixture of ferromagnetic and antiferromagnetic features, a necessary condition for the occurrence of spin-glasses.

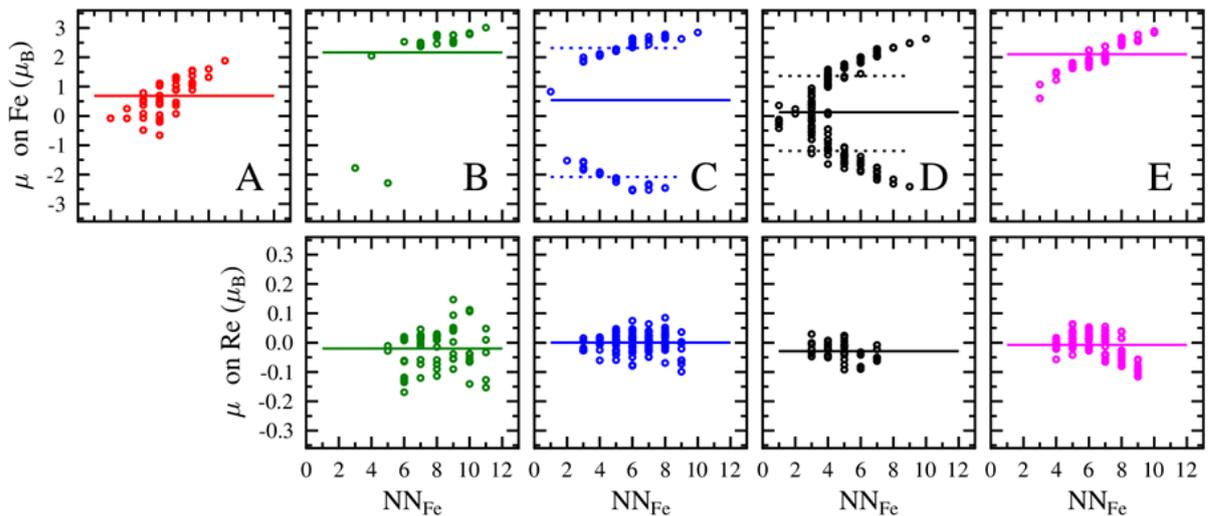

Fig. 12

(Color online) Theoretically calculated magnetic moments, using the APM model, on Fe and Re atoms occupying the five different sub lattices A, B, C, D and E versus number of nearest-neighbor Fe atoms, $NN_{Fe}$. The cloud of data seen for each sub lattice represent different atomic configurations taken into account. Solid lines indicate the average values of the magnetic moments, whereas the average values for the positive and negative clouds are marked by dotted lines.

### 4.3.1. KKR-CPA calculations

The aforementioned KKR results can be compared to the KKR-CPA calculations, which partially accounted for a random distribution of Fe and Re in the system. Chemical disorder was treated on B, C and either on D or E crystallographic sites, whereas the A position was assumed to be fully occupied by Fe atoms to simplify computations (In fact, according to experimental finding the population of Fe atoms on A sites was $\geq 90\%$ [8]). Otherwise, in the case of binary alloys with atoms distributed over thirty sites, the self-consistent CPA cycle would require computing $2^{30}$ (more than $10^9$) configurations, a task numerically impossible to do. The number of configurations reasonably decreases to $2^{20}$ (~$10^6$) or even less viz. $2^{16}$=65536 configurations in the considered cases. In this partially disordered model, we have calculated a spin-polarized density of states and local magnetic moments on Fe and Re sites for $Fe_{17}Re_{13}$ and $Fe_{15}Re_{15}$ compounds considering two types of magnetic arrangements (FM and APM).

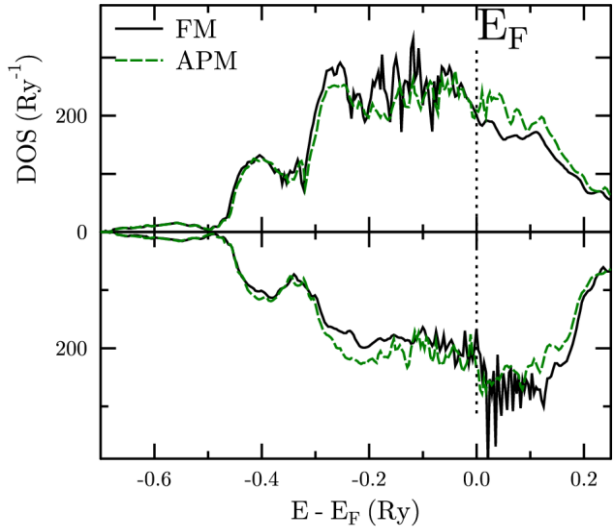

Fig. 13

(Color online) Total electronic DOS in the $\sigma$-$Fe_{15}Re_{15}$ alloy obtained with the FM (black, solid) and APM (green, dashed) model, as calculated by the KKR-CPA method.

The first important difference of the KKR-CPA results with respect to the averaged KKR results was a slightly smaller total magnetic moment (per atom) calculated both for FM and APM models. This is mainly due to almost vanishing Fe moment on the site A as well as to a smaller value of the moment on the site E. We suppose that the CPA approach, which takes into account all possible local atomic arrangements when computing DOS and magnetic moments, is more accurate to describe electronic and magnetic properties of such highly disordered Fe-Re sigma-phase. Thus, one can expect that the results obtained from (i) averaging of finite number of ordered models (the KKR approach), and (ii) disordered model (the KKR-CPA approach), can be slightly different. In the case of other lattice sites, the agreement between both approaches is really fine.

Noteworthy, the magnetic moment for $Fe_{15}Re_{15}$ ($\mu = 0.29$ $\mu_B$) as found with the KKR-CPA calculations is in very close agreement with the corresponding experimental value (~0.30 $\mu_B$). It can be also seen (Table I) that the average value slightly increases with increasing Fe concentration, which also fits to the experimental data.

It is also interesting to note that from both KKR and KKR-CPA computations, and also for both considered models of the magnetic ordering, large values of Fe magnetic moments were obtained, whatever the crystallographic sites (except for the less populated A site). Further, the lowering of the total magnetic moment can be achieved only by a relative orientation of the Fe-site magnetic moments (locally anti-parallel), without substantial decrease of their absolute values. The magnetic behavior can be best seen on total and site-decomposed DOS calculated in FM and APM states (Figs 13 and 14) for the $Fe_{15}Re_{15}$ compound.

**Table I.** Total and Fe magnetic moments (in $\mu_B$ per atom) on the five crystallographic sites in Fe-Re sigma phase as computed with the KKR-CPA calculations both for FM and APM model.

|  | Total | A-site | B-site | C-site | D-site | E-site |
|---|---|---|---|---|---|---|
| FM | | | | | | |
| $Fe_{17}Re_{13}$ | 0.93 | -0.13 | 2.49 | 2.67 | 1.65 | 1.52 |
| $Fe_{15}Re_{15}$ | 0.78 | -0.36 | 2.68 | 2.66 | 1.63 | 1.74 |
| APM | | | | | | |
| $Fe_{17}Re_{13}$ | 0.32 | -0.15 | 2.72 | -2.20/2.40 | -1.31/1.54 | 1.61 |
| $Fe_{15}Re_{15}$ | 0.29 | -0.10 | 2.52 | -2.33/+2.34 | -1.49/1.55 | 1.72 |

For example, DOS of Fe on B and C sites, which exhibit the largest spin-polarization, look very similar for both magnetic ordering models. Noteworthy, in the case when the APM ordering was assumed (C and D sites) two types of DOS, of almost anti-symmetric shape, were obtained for Fe atoms (also for Re atoms, but their spin-polarization is actually negligible), represented by corresponding indices ($Fe_u$ or $Fe_d$). In Fig. 14 we show the corresponding DOS (on C and D sites) only for the case of $Fe_u$. Interestingly, the similar shape of DOS was also detected for the remaining three sites (A, D and E) where the spin-polarization is smaller. The finding clearly shows that the absolute values of Fe magnetic moments, in so strongly disordered system, are slightly affected by their relative directions.

## 4. 4. Itinerant character of magnetism

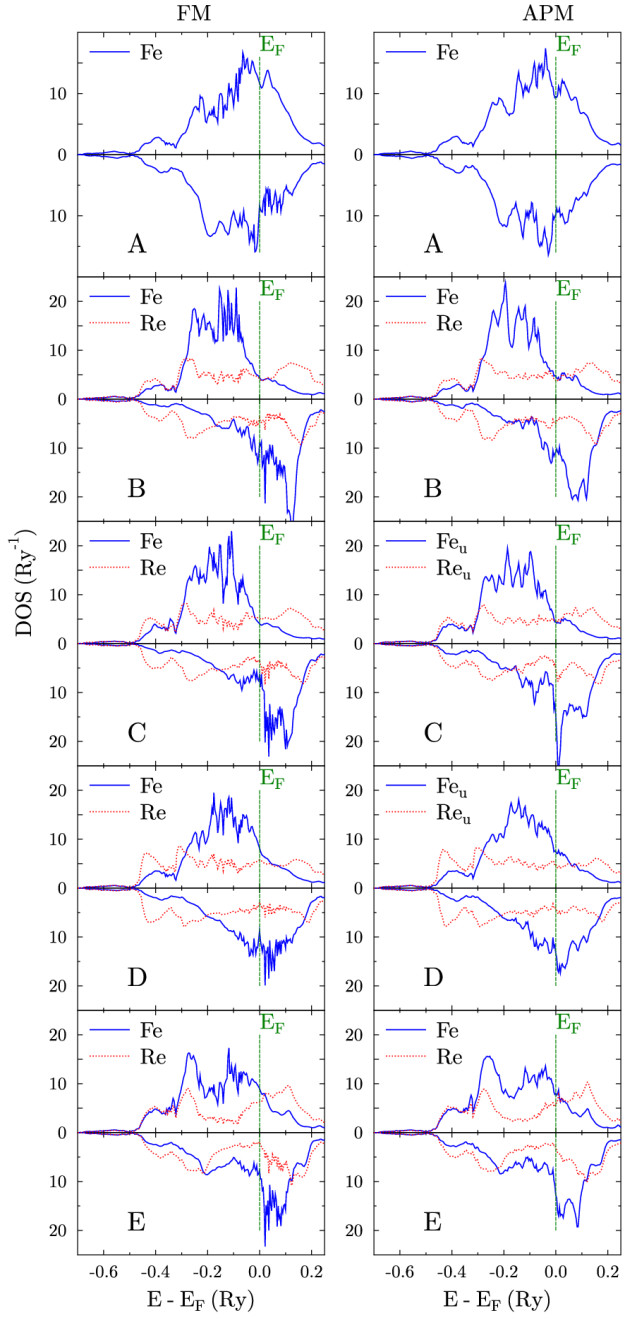

Based on the magnetization measurements the character of the magnetism in the studied compounds can be determined. For that purpose one can use the Rhodes-Wohlfarth criterion [20] according to which a magnetism exhibits an itinerant or localized character depending on the value of a ratio between the effective magnetic moment in a paramagnetic phase, $\mu_{eff}$, and that in the ordered phase, $\mu$. If the ratio is higher than 1 than the magnetism is itinerant, otherwise it is localized. The values of $\mu_{eff}/\mu$ calculated for the studied compounds are shown in Fig. 15 which illustrates the figure of merit for several chosen localized and itinerant magnets in form of the so-called Rhodes-Wohlfarth plot. The plot clearly demonstrates that the magnetism of the σ-phase Fe-Re compounds has a highly itinerant character. Although the latter weakens with the increase of the content of iron, yet it remains strongly itinerant even for the iron-most concentrated sample.

Fig. 14
(Color online) Site-decomposed KKR-CPA DOS for the sigma-$Fe_{15}Re_{15}$ alloy as obtained with the APM model.

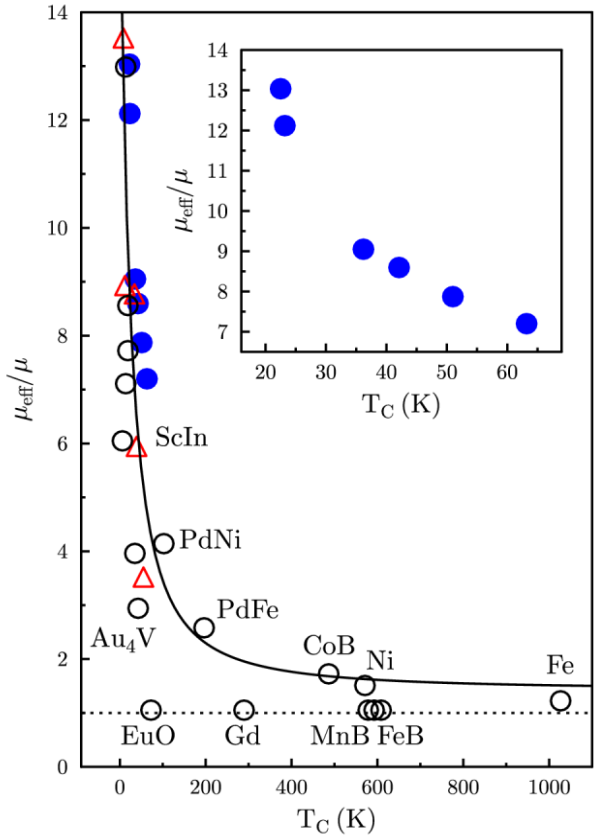

Fig. 15

(Color online) Rhodes-Wohlfarth plot for various itinerant and localized magnetic systems (open circles) including the σ-FeCr compounds (open triangles) [4], and presently investigated samples (full circles). The latter are also shown as inset. The horizontal dotted line indicates the critical value above which the magnetism is itinerant.

## 5. Conclusions

Magnetism in a series of σ-Fe$_{100-x}$Re$_x$ intermetallic compounds with *x* ranging between 43 and 53 was revealed by means of experimental measurements and theoretical calculations. The results obtained in this study can be summarized by drawing the following conclusions:

1. All investigated samples, covering the whole range of the sigma-phase occurrence, exhibit magnetism: the ordering temperature changes monotonically between ~65K for σ-Fe$_{53}$Re$_{47}$ and ~23K for σ-Fe$_{47}$Re$_{53}$.

2. The magnetism has a spin-glass phase as the ground state, possibly with a re-entrant character and a ferromagnetic-like phase as an intermediate state: the spin-freezing temperature decreases in a monotonic way with the increase of rhenium content from ~50 K to ~18 K for the limiting concentrations.

3. The spin-glass state can be subdivided into two domains characterized by: (a) a weak irreversibility, and (b) a strong irreversibility, separated by a peak in the ZFC curve.

4. The degree of irreversibility strongly grows with a decrease of temperature for all samples. Its temperature-integrated value hardly depends on Re-content for 43 ≤ *x* ≤ 49, and is significantly lower for *x*≈53.

5. The magnetism has a highly itinerant character, the degree of which decreases with the increase of iron concentration.

6. Theoretical calculations admitting a co-linear ferromagnetic ordering yielded overestimated, by a factor of ~2.5, values of the average magnetic moment, while those admitting a co-linear but antiparallel ordering on two of five sub lattices resulted in a significant improvement.

7. Theoretically calculated absolute values of Fe magnetic moments are high in all applied calculation methods and models, they are also slightly affected by their relative directions. Lowering the average magnetic moment can be achieved by applying various orientations of the partial moments located on the various lattice sites.

**Acknowledgements**

The Ministry of Science and Higher Education, Warszawa, is thanked for a support.